\documentclass{ws-procs9x6}
\newcommand\be{\begin{equation}}
\newcommand\ee{\end{equation}}
\newcommand\ba{\begin{eqnarray}}
\newcommand\ea{\end{eqnarray}}
\newcommand\eq{\begin{equation}}
\newcommand\en{\end{equation}}

\begin{document}

\title{Inflation model building in moduli space\footnote{\uppercase{B}ased on the talk presented by \uppercase{KK} at \uppercase{PASCOS04} workshop.}}

\author{Kenji Kadota\footnote{\uppercase{W}ork partially
supported by \uppercase{NASA} grant \uppercase{NAG5-10842.}}}

\address{NASA/Fermilab Theoretical Astrophysics Center, \\
Fermilab, Batavia, IL 60510, USA
%\\ E-mail: kadota@fnal.gov
}

\author{Ewan D. Stewart\footnote{\uppercase{W}ork partially
supported by \uppercase{A}strophysical \uppercase{R}esearch \uppercase{C}enter for the \uppercase{S}tructure and \uppercase{E}volution
of the \uppercase{C}osmos funded by the \uppercase{K}orea \uppercase{S}cience and \uppercase{E}ngineering \uppercase{F}oundation
and the \uppercase{K}orean \uppercase{M}inistry of \uppercase{S}cience,
and by the \uppercase{K}orea \uppercase{R}esearch \uppercase{F}oundation grants \uppercase{KRF-2000-015-DP0080} and \uppercase{KRF PBRG
2002-070-C00022}.}}

\address{Department of Physics, KAIST, Daejeon 305-701, South Korea 
% \\E-mail: wspc@wspc.ox.uk
}  

\maketitle
\abstracts{
A self-consistent modular cosmology scenario and its testability in view of future CMB experiments 
are discussed. Particular attention is drawn to the enhanced symmetric points in moduli space which play crucial roles in our scenario.
The running and moreover the running of running for the cosmic perturbation spectrum are also analyzed. 
}
\section{Inflation model building}
\label{sprit}
The important questions we would like to answerer in inflation model building are
\begin{center}
{\em What is the inflaton field?} \\
{\em What are its properties?} 
\end{center}

To try to answer these questions, we need to consider what is a natural particle theory framework to discuss the dynamics of the early Universe.
\\
(i){\em What fields were there in the early Universe?}
\\
 The most promising candidate to describe the physics of the early universe is string theory, and string theory has many flat directions whose potentials vanish in exact supersymmetry. The fields parameterizing such flat directions are called moduli, and we shall 
discuss if a moduli field can realize a successful inflationary scenario or not. 
\\
(ii) {\em What does a modulus potential look like?}  
\\
The properties of moduli fields are heavily dependent on the way supersymmetry is broken. In the following, we discuss a hidden sector symmetry breaking scenario, in which the generic form of the moduli potential becomes \ba
\label{modularpot}
V(\phi)=M_{s}^4F\left(\frac{\phi}{M_p}\right)
\ea
where $M_s$ is the supersymmetry breaking scale and $F$ is a dimensionless 
function with of order unity coefficients.  
In particular, for gravity mediated supersymmetry breaking, $M_s\approx 10^{10-11}$GeV and 
the mass of modulus $\phi$ consequently becomes $m_{\phi}=M_{s}^2/M_p \approx 10^{2-3} $GeV.
Note, as is usually the case for supergravity inflation, the slow-roll parameter $V''/V$ is of order unity, which is a generic problem for supergravity inflation unless we choose a special form of K\"ahler potential.
  
\section{Can a consistent inflationary scenario be realized in this natural context?}
The detailed discussion of a possible self-consistent modular cosmology scenario based on this natural simple particle theory setup 
is given in 
%Ref.~\refcite{suc}
[\refcite{suc}]. We shall concentrate on the predicted cosmic perturbations for this scenario in this article. Note that we will consider a single modulus field which is complex. This does not mean we add an additional degree of freedom (namely angular component in addition to the radial component) because all scalar fields are complex in supersymmetry. 
The overview of the dynamics of the inflaton modulus is the following. 

Near a maximum, the potential of Eq.(\ref{modularpot}) has the form\footnote{This form relies on the maximum being a point of enhanced symmetry. See [\refcite{pert}] for details.}
\be
V(\phi) = V_0 - \frac{1}{2} m_\phi^2 |\Phi|^2 + \ldots~,
\ee
where $V_0 \sim M_\mathrm{s}^4$ and $m_\phi \sim M_\mathrm{s}^2$.   
Our scenario starts with an eternally inflating universe consisting of an ensemble of eternally inflating extrema throughout the field space of string theory, which avoids the Brustein-Steinhardt problem \cite{bs}. Our local universe is a region where the field rolled down from its maximum to escape from the eternally inflating region, and the observable cosmic perturbations are produced while the field rolls down to its minimum. For a single complex modulus
\be
\Phi = \frac{1}{\sqrt{2}\,} \phi e^{i \theta}~,
\ee
the total curvature perturbations arise from radial and angular component fluctuations \cite{deltaN} 
\be
\mathcal{R}_\mathrm{c} = \delta N
= \frac{\partial N}{\partial \phi} \, \delta\phi
+ \frac{\partial N}{\partial \theta} \, \delta\theta ~.
\ee

Before presenting the detailed results in Sec \ref{obs}, simple estimates tell us that the contribution of radial component fluctuations to the final curvature perturbations is
\be
\label{rad2}
\frac{\partial N}{\partial \phi} \, \delta\phi 
= - \frac{H}{\dot{\phi}} \, \delta\phi
\sim \frac{H}{\phi}~,
\ee
and that for the angular component is
\be\label{ang2}
\frac{\partial N}{\partial \theta} \, \delta\theta 
\sim \frac{\partial N}{\partial \theta} \frac{H}{\phi}~.
\ee
A crucial point here is that the angular component perturbations dominate for a large enough $\frac{\partial N}{\partial \theta}$.

\subsection{Points of enhanced symmetry}
The existence of points of enhanced symmetry is a robust and unique feature of moduli space. At those special points, some fields (matter, gauge, or moduli) become massless 
and become massive away from it, being Higgsed by the modulus that parameterizes the distance from the point of enhanced symmetry. 
At such a point of enhanced symmetry, the couplings of the modulus to those light degrees of freedom 
will cause the moduli mass to renormalize as a function of
$\phi$.
This can turn the mass squared of the modulus positive for small
$\phi$, shifting the maximum of the potential out to some finite value
$\phi = \phi_*$. The modulus maximum now becomes a rim and the field starts rolling down from there. The expression for the angular component fluctuation given in Eq.(\ref{ang2})
had a steep spectrum because $\phi$ in the expression $H/\phi$ changes rapidly due to non-slow-roll. It now, taking account of the loop corrections in the potential, goes like $ H/\phi_0$ where $\phi_0$ is radius of the rim maximum of the potential, 
making the spectrum flat. The remaining problem is if large ${\partial N}/ \partial \theta$ is probable, or, if the initial angle $\theta$ corresponding to large ${\partial N}/ \partial \theta$ is probable. It is indeed probable, if we consider that the regions corresponding to those initial angles leading to large ${\partial N}/ \partial \theta$ would undergo greater
expansion and hence occupy a larger volume at late times.

 It turns out, as discussed in detail in [\refcite{pert}], that the renormalized potential induced by the effects of light degrees of freedom at the saddle points identified as enhanced symmetric points will dynamically 
select the desirable initial angle for the inflaton modulus.

\subsection{Observable predictions}
\label{obs}
In [\refcite{pert}], we performed detailed analytic and numerical calculations for our modular cosmology scenario without assuming slow-roll conditions. The perturbation spectrum in our model has 
negligible deviations from scale invariance over a wide range of $k$ with running becoming significant at (very) small (but possibly observable) scales.
The form of the spectral index turned out to be a simple polynomial form
\begin{equation}\label{parame}
n-1 = A k^\alpha
\end{equation}
Its running and running of running
\begin{equation}
\frac{dn}{d\ln k} = \alpha A k^\alpha,~~\frac{d^2n}{(d\ln k)^2} = \alpha^2 A k^\alpha
\end{equation}
illustrate our theoretical expectations that the usually assumed hierarchy $|n''| \ll |n'| \ll |n-1| \ll 1$ is valid 
only for a limiting region of parameter space, $\alpha \ll 1$, where the running would in any 
case be small, while for a wider range of $\alpha$ the running is significant but so is the running of the running, $|n''| \sim |n'| \sim |n-1|$.

It is also worth noting that Eq.~(\ref{parame}) indicates that the running and the running of running are expected to be 
most significant toward smaller scales, \mbox{i.e.} negligible $n'$ at large 
scales does not necessarily guarantee the absence of running in the spectrum, which cannot be taken into account if we ignore $n''$.
Thus it is crucial for our observations to probe the smallest possible scales to search for a signal of running.

\section{Discussion}
There are several works which 
try to explain the apparent discrepancy between the natural energy scale of the moduli potential 
($\sim 10^{10-11}$GeV) and the energy scale of inflation ($\sim 10^{15}$GeV) obtained in some simple inflation models. They tend to try to find a non-trivial mechanism to scale up the energy scale of the modulus potential to the GUT scale to match the amplitude of fluctuations with observations. We, however, instead stuck with the value of 
$\sim 10^{10-11} $GeV and saw if our natural particle theory setup leads to the observationally consistent inflationary scenario. We discussed the cosmic perturbations in our scenario, which gives one of the most stringent constraints on an inflation model, with a particular emphasis on the possible running of running which in general cannot be ignored for the consistency of perturbation calculations. 
%in general cannot be ignored for the consistency of perturbation calculations 
%We discussed the significance of enhanced symmetric points in moduli space which plays crucial roles to obtain the flat perturbation spectrum as 
%well as the dynamical selection of desired initial angle for inflaton modulus without fine-tuning. 
%In performing the detailed cosmic perturbation calculations without 
%assuming the slow-roll approximations, we emphasized the possible running of running which 
%in general cannot be ignored for the consistency of perturbation calculations.
%If one expects the running would more likely show up for the scales which exit the horizon toward the end of inflation, 
%the low energy scale inflation would be preferable for the detection of running, because scales 
%CMB can probe exit the horizon closer to the end of inflation for smaller inflation energy scale.
%If this is the case, observing both the running of spectrum and the gravitational 
%wave signatures would be less likely (if CMB can detect one of them at all) because the high 
%inflation energy scale is preferable for the gravitational wave detection.

In addition to the problems discussed in this article, another well-known and long-standing problem is the cosmological moduli problem. We proposed a baryogenesis scenario following thermal inflation in [\refcite{ad}] to make this aspect of modular cosmology self-consistent too. Besides these phenomenological aspects of modular cosmology, more fundamental 
problems, such as moduli stabilization, are also under intense investigation \cite{string}. 
The ubiquitous existence of moduli is a generic prediction of string/M-theory, and the realization of a successful modular cosmology 
scenario would be worth further study.

%\section*{Acknowledgments}
%KK thanks PASCOS04 organizers for the fruitful and stimulating workshop.

\end{document}